\documentstyle[preprint,aps]{revtex}

\begin{document}

\draft
\tighten
\topmargin-48pt
\evensidemargin5mm
\oddsidemargin5mm

\preprint{EFUAZ FT-95-16-REV}

\title{About the Claimed `Longitudity' of the Antisymmetric
Tensor Field After Quantization\thanks{Submitted to ``Physical Review D".}}

\author{{\bf Valeri V. Dvoeglazov}\thanks{On leave of absence from
{\it Dept. Theor. \& Nucl. Phys., Saratov State University,
Astrakhanskaya ul., 83, Saratov\, RUSSIA.} Internet
address: dvoeglazov@main1.jinr.dubna.su}}

\address{
Escuela de F\'{\i}sica, Universidad Aut\'onoma de Zacatecas \\
Antonio Doval\'{\i} Jaime\, s/n, Zacatecas 98068, ZAC., M\'exico\\
Internet address:  VALERI@CANTERA.REDUAZ.MX
}

\date{January 20, 1996. First version: July, 1995}

\maketitle

\begin{abstract}
It has long been claimed that antisymmetric tensor field of the second
rank is longitudinal after quantization. Such a situation is quite
unacceptable from a viewpoint of the Correspondence Principle. On the
basis of the Lagrangian formalism we calculate the Pauli-Lyuban'sky vector
of relativistic spin for this field.  Even at the classical level it can
be equal to zero after application of the well-known constraints.  The
correct quantization procedure  permits us to propose solution of this
puzzle in the modern field theory.   Obtained results develop the previous
consideration of Evans [{\it Physica A}214 (1995) 605-618].
\end{abstract}


\newpage

\baselineskip13pt

Quantum electrodynamics (QED) is a construct which found overwhelming
experimental confirmations (for recent reviews see, {\it e.g.},
refs.~\cite{BS1,BS2}). Nevertheless, a number of theoretical aspects
of this theory deserves more attention. First of all, they are:
the problem of ``fictious photons of helicity other than $\pm j$, as well
as the indefinite metric that must accompany them"; the renormalization
idea, which ``would be sensible only if it was applied with finite
renormalization factors, not infinite ones (one is not allowed to neglect
[and to subtract] infinitely large quantities)"; contradictions with the
Weinberg theorem ``that no symmetric tensor field of rank $j$ can be
constructed from the creation and annihilation operators of massless
particles of spin $j$",\, {\it etc.} They were shown at by
Dirac~\cite{Dirac1,Dirac2} and by Weinberg~\cite{Weinberg}.  Moreover,
it appears now that we do not yet understand many specific peculiarities
of classical electromagnetism, first of all, the  problems of longitudinal
modes and of the Coulomb action-at-a-distance,
refs.~\cite{Evans,Evans1,Staru,DVO1,DVO2,DVO3,DVO4,Chubykalo}.  Secondly,
the standard model, which has been constructed on the base of ideas,
which are similar to QED, appears to be no able to explain many
puzzles in neutrino physics.

In my opinion, all these shortcomings are the consequences
of ignoring several important questions.
``In the classical electrodynamics of charged particles, a
knowledge of $F^{\mu\nu}$ completely determines the properties of the
system. A knowledge of $A^\mu$ is redundant there, because it is
determined only up to gauge transformations, which do not affect
$F^{\mu\nu}$\ldots  Such is not the case in quantum
theory\ldots"~\cite{Huang}. We learnt, indeed, about this fact from
the Aharonov-Bohm~\cite{A1} and the Aharonov-Casher effects~\cite{A2}.
However, recently several attempts have been undertaken to explain the
Aharonov-Bohm effect classically~\cite{A3}.
These attempts have, in my opinion, logical basis and are in complete
accordance with the Correspondence Principle.  In the mean time,
quantization of the antisymmetric tensor field led us to a new puzzle,
which until now was not drawn much attention to.  It was claimed that the
antisymmetric tensor field of the second rank is longitudinal after
quantization~\cite{Ogievet,Hayashi,Love,AVD,Sorella}. We know that the
antisymmetric tensor field (electric and magnetic fields, indeed) is
transversal in the Maxwellian classical electrodynamics. It is clear that
longitudinal components can not be transformed into the transversal ones
in the $\hbar \rightarrow 0$ limit.\footnote{See also group-theoretical
consideration in ref.~\cite{Kim} which concerns with the reduction of
transversal rotational degrees of freedom to gauge degrees of freedom in
infinite-momentum/zero-mass limit.  The only mentions of the
transversality of the quantized antisymmetric tensor field see in
refs.~\cite{Takahashi,Boyarkin}.} How should we manage with the
Correspondence Principle in this case? It is often concluded:  one is not
allowed to use the antisymmetric tensor field to represent the quantized
electromagnetic field in relativistic quantum mechanics.  Nevertheless, we
are convinced that a reliable theory should be constructed on the base of
a minimal number of ingredients (``Occam's Razor") and should have
well-defined classical limit.  Therefore, in this paper we undertake a
detailed analysis of rotational properties of the antisymmetric tensor
field, we calculate the Pauli-Lyuban'sky vector of relativistic spin
(which defines, what the field is: transversal or longitudinal) and we
then conclude, whether it is possible to obtain conventional
electromagnetic theory with transversal modes provided that strengths
({\it not} potentials) are chosen to be physical variables.  The
particular case also exists when the Pauli-Lyuban'sky vector for the
antisymmetric tensor field of the second rank is equal to zero, what
corresponds to the claimed longitudity (helicity $h=0$) of this field.

Researches in this area from a viewpoint of the Weinberg's $2(2j+1)$
component theory have been started in
refs.~\cite{DVA00,DVA0,DVO00,DVO01,DVO02,DVO1,DVO2,DVO3,DVO4}.
I would also like to point out that the problem at hand is directly
connected with our understanding of the nature of neutral particles,
including neutrinos~\cite{Majorana,MLC,Ziino,DVA1,DVA2,DVO5,DVO6,DVO7}.
From a mathematical viewpoint theoretical content does not depend,
what representation space, which field operators transform on, has been
chosen.

We begin with the antisymmetric tensor field operator (in general,
complex-valued):
\begin{equation}
F^{\mu\nu} (x)
\,=\, \sum_\eta \int \frac{d^3 {\bf p}}{(2\pi)^3} \, {1\over 2E_p}\, \left
[ F^{\mu\nu}_{\eta\,(+)} ({\bf p})\, a_\eta ({\bf p})\, e^{-ip\cdot x} +
F^{\mu\nu}_{\eta\, (-)} ({\bf p}) \,b_\eta^\dagger ({\bf p})\, e^{+ip\cdot
x} \right ]\label{fop}
\end{equation}
and with the Lagrangian, including, in general, mass term:\footnote{The
massless limit ($m\rightarrow 0$) of the
Lagrangian is connected with the Lagrangians used in the conformal field
theory and in the conformal supergravity by adding the total derivative:
\begin{equation}
{\cal L}_{CFT} = {\cal L} + {1\over 2}\partial_\mu \left
( F_{\nu\alpha} \partial^\nu F^{\mu\alpha} - F^{\mu\alpha} \partial^\nu
F_{\nu\alpha} \right )\quad.
\end{equation}
The gauge-invariant
form~\cite{Ogievet} is obtained only if take into account the generalized
Lorentz condition, see ref.~\cite{Hayashi} and what is below.}
\begin{equation}\label{Lagran}
{\cal L} =  {1\over 4} (\partial_\mu
F_{\nu\alpha})(\partial^\mu F^{\nu\alpha}) - {1\over 2} (\partial_\mu
F^{\mu\alpha})(\partial^\nu F_{\nu\alpha}) - {1\over 2} (\partial_\mu
F_{\nu\alpha})(\partial^\nu F^{\mu\alpha}) + {1\over 4} m^2 F_{\mu\nu}
F^{\mu\nu} \quad.
\end{equation}
The Lagrangian leads to the equation of motion in the following form:
\begin{equation}
{1\over 2}
({\,\lower0.9pt\vbox{\hrule \hbox{\vrule height 0.2 cm \hskip 0.2 cm
\vrule height 0.2cm}\hrule}\,}+m^2) F_{\mu\nu} +
(\partial_{\mu}F_{\alpha\nu}^{\quad,\alpha} -
\partial_{\nu}F_{\alpha\mu}^{\quad,\alpha}) = 0 \quad,\label{PE}
\end{equation}
where ${\,\lower0.9pt\vbox{\hrule \hbox{\vrule height 0.2 cm
\hskip 0.2 cm
\vrule height 0.2 cm}\hrule}\,}
=- \partial_{\alpha}\partial^{\alpha}$.
It is this equation for antisymmetric-tensor-field components
that follows from the Proca-Bargmann-Wigner consideration:
\begin{eqnarray}\label{Proca}
&&\partial_\alpha F^{\alpha\mu} + {m\over 2} A^\mu = 0 \quad, \\
&&2 m F^{\mu\nu} = \partial^\mu A^\nu - \partial^\nu A^\mu \quad,
\end{eqnarray}
provided that $m\neq 0$ and in the final expression we take into account
the Klein-Gordon equation $({\,\lower0.9pt\vbox{\hrule \hbox{\vrule height
0.2 cm \hskip 0.2 cm \vrule height 0.2 cm}\hrule}\,} - m^2) F_{\mu\nu}=
0$. The latter expresses relativistic dispersion relations $E^2 -{\bf p}^2
=m^2$ and it follows from the coordinate Lorentz transformation
laws~\cite[\S 2.3]{Ryder}.

Following the variation procedure given, {\it e.g.}, in
refs.~\cite{Corson,Barut,Bogoliubov} one can obtain that
for rotations $x^{\mu^\prime} = x^\mu + \omega^{\mu\nu} x_\nu$
the corresponding variation of the wave function is found
from the formula:
\begin{equation}
\delta F^{\alpha\beta} = {1\over 2} \omega^{\kappa\tau}
{\cal T}_{\kappa\tau}^{\alpha\beta,\mu\nu} F_{\mu\nu}\quad.
\end{equation}
The generators of infinitesimal transformations are then defined as
\begin{eqnarray}
\lefteqn{{\cal T}_{\kappa\tau}^{\alpha\beta,\mu\nu} \,=\,
{1\over 2} g^{\alpha\mu} (\delta_\kappa^\beta \,\delta_\tau^\nu \,-\,
\delta_\tau^\beta\,\delta_\kappa^\nu) \,+\,{1\over 2} g^{\beta\mu}
(\delta_\kappa^\nu\delta_\tau^\alpha  \,-\,
\delta_\tau^\nu\, \delta_\kappa^\alpha) +\nonumber}\\
&+&\,
{1\over 2} g^{\alpha\nu} (\delta_\kappa^\mu \, \delta_\tau^\beta \,-\,
\delta_\tau^\mu \,\delta_\kappa^\beta) \,+\, {1\over 2}
g^{\beta\nu} (\delta_\kappa^\alpha \,\delta_\tau^\mu \,-\,
\delta_\tau^\alpha \, \delta_\kappa^\mu)\quad.
\end{eqnarray}
It is ${\cal T}_{\kappa\tau}^{\alpha\beta,\mu\nu}$, the generators of
infinitesimal transformations,
that enter in the formula for the relativistic spin tensor:
\begin{equation}
J_{\kappa\tau} = \int d^3 {\bf x} \left [ \frac{\partial {\cal
L}}{\partial ( \partial F^{\alpha\beta}/\partial t )} {\cal
T}^{\alpha\beta,\mu\nu}_{\kappa\tau} F_{\mu\nu} \right ]\quad.
\label{inv}
\end{equation}
As a result we obtain:
\begin{eqnarray}
J_{\kappa\tau} &=& \int d^3 {\bf x} \left [ (\partial_\mu F^{\mu\nu})
(g_{0\kappa} F_{\nu\tau} - g_{0\tau} F_{\nu\kappa}) -  (\partial_\mu
F^\mu_{\,\,\,\,\kappa}) F_{0\tau} + (\partial_\mu F^\mu_{\,\,\,\,\tau})
F_{0\kappa} + \right. \nonumber\\
&+& \left. F^\mu_{\,\,\,\,\kappa} ( \partial_0 F_{\tau\mu} +
\partial_\mu F_{0\tau} +\partial_\tau F_{\mu 0})  -   F^\mu_{\,\,\,\,\tau}
( \partial_0 F_{\kappa\mu} +\partial_\mu F_{0\kappa} +\partial_\kappa
F_{\mu 0}) \right ]\quad. \label{gene}
\end{eqnarray}
If agree that the
orbital part of the angular momentum \begin{equation} L_{\kappa\tau} =
x_\kappa \Theta_{0\,\tau} - x_\tau \Theta_{0\,\kappa} \quad,
\end{equation}
with  $\Theta_{\tau\lambda}$ being the energy-momentum tensor, does not
contribute to the Pauli-Lyuban'sky operator when acting on the
one-particle free states (like the Dirac $j=1/2$ case), then
the Pauli-Lyuban'sky 4-vector is constructed as
follows~\cite[Eq.(2-21)]{Itzykson}
\begin{equation}
W_\mu = -{1\over 2}  \epsilon_{\mu\kappa\tau\nu} J^{\kappa\tau} P^\nu \quad,
\end{equation}
with $J^{\kappa\tau}$ defined by Eqs.
(\ref{inv},\ref{gene}). The 4-momentum opertor $P^\nu$ can be replaced by
its eigenvalue when acting on plane-wave eigenstates. Then we use the
Lagrangian (\ref{Lagran}) and choose space-like normalized vector $n^\mu
n_\mu = -1$ (for example $n_0 =0$,\, ${\bf n} = \widehat  {\bf p} = {\bf
p} /\vert {\bf p}\vert$\, ).  After lengthy calculations in a spirit
of~\cite[p.58,147]{Itzykson} one can find the explicit form of the
relativistic spin:\footnote{Let me remind that the helicity operator is
connected with the Pauli-Lyuban'sky vector in the following manner $({\bf
J} \cdot \widehat {\bf p}) = ({\bf W} \cdot \widehat {\bf p})/ E_p$, see
ref.~\cite{Shirok}.}
\begin{mathletters}
\begin{eqnarray}
&& (W_\mu \cdot n^\mu) = - ({\bf W}\cdot {\bf n}) = -{1\over 2}
\epsilon^{ijk} n^k J^{ij} p^0\quad,\label{PL1}\\
&& {\bf J}^k = {1\over 2}
\epsilon^{ijk} J^{ij} = \epsilon^{ijk} \int d^3 {\bf x} \left [ F^{0i}
(\partial_\mu F^{\mu j}) + F_\mu^{\,\,\,\,j} (\partial^0 F^{\mu i}
+\partial^\mu F^{i0} +\partial^i F^{0\mu} ) \right ]\quad.\label{PL2}
\end{eqnarray}
\end{mathletters}
Now it becomes clear that application of the generalized Lorentz
conditions (which are quantum versions of free-space dual Maxwell's
equations) leads in such a formulation to the absence of electromagnetism
in a conventional sense.\footnote{I would still like to point out one paradox
connected with the presented treatment and the ordinary electrodynamics.
On the basis of definitions:
\begin{equation}
\widetilde F^{\alpha\beta}
={1\over 2} \epsilon^{\alpha\beta\gamma\delta} F_{\gamma\delta}\quad,\quad
F^{\alpha\beta} = - {1\over 2} \epsilon^{\alpha\beta\gamma\delta}
\widetilde F_{\gamma\delta}
\end{equation}
and the use of the mathematical theorem that {\it any} antisymmetric
tensor of the second rank can be expanded in potentials,
{\it e.g.}, ref.~\cite{Bongaarts} in the following manner:
\begin{equation}
F_{\alpha\beta} = \partial_\alpha A_\beta -
\partial_\beta A_\alpha\quad,\quad \widetilde F_{\alpha\beta} =
\partial_\alpha \tilde A_\beta - \partial_\beta \tilde A_\alpha
\end{equation}
one can deduce the equivalent relations:
\begin{mathletters}
\begin{eqnarray}
\partial_\nu F_{\alpha\beta} +
\partial_\alpha F_{\beta\nu} +\partial_\beta F_{\nu\alpha} = 0 \,
&\Leftrightarrow& \, \epsilon^{\mu\nu\alpha\beta} \partial_\nu
F_{\alpha\beta} = 0 \,\, \Leftrightarrow \,\, \partial_\nu \widetilde
F^{\mu\nu} = 0 \quad,\\ \partial_\nu \widetilde F_{\alpha\beta} +
\partial_\alpha \widetilde F_{\beta\nu} +\partial_\beta \widetilde
F_{\nu\alpha} = 0 \, &\Leftrightarrow& \, \epsilon^{\mu\nu\alpha\beta}
\partial_\nu \widetilde F_{\alpha\beta} = 0 \,\, \Leftrightarrow \,\,
\partial_\nu F^{\mu\nu} = 0\quad.
\end{eqnarray}
\end{mathletters}
These relations state that currents cannot be put in the Maxwell's
equations. Therefore, something should be corrected in our understanding
of the nature of dual second-rank antisymmetric tensors, 4-currents and/or
in the mentioned theorem.} The resulting Kalb-Ramond field is longitudinal
(helicity $h=0$).  The discussion of this fact can also be found in
ref.~\cite{Hayashi,DVO2}.

One of possible ways to  obtain helicities $h=\pm 1$
is a modification of the electromagnetic field tensor like ref.~[6q],
{\it i.e.}, introducing the non-Abelian electrodynamics~\cite{Evans1}:
\begin{equation}
F_{\mu\nu}\quad \Rightarrow \quad
{\bf G}_{\mu\nu}^{a} = \partial_\mu A_\nu^{(a)\, \ast}
- \partial_\nu A_\mu^{(a)\, \ast} -i{e\over \hbar}[ A_\mu^{(b)},
A_\nu^{(c)}] \quad,
\end{equation}
where $(a),\,(b),\,(c)$ are the vector components in the $(1),\,(2),\,(3)$
circular basis~\cite{Evans,Evans1}. In the other words, one can add some
ghost field (the $B(3)$ field) to the antisymmetric  tensor $F_{\mu\nu}$.
As a matter of fact this induces hypotheses on a massive photon
and/or an additional displacement current. We prefer to avoid any
auxiliary constructions (even they are valuable in intuitive explanations
and generalizations). If these constructions exist they should be deduced
from a more general theory on the basis of some fundamental  postulates,
{\it e.g.}, in a spirit of refs.~\cite{DVA0,DVO95}.  In the procedure of
quantization  one can reveal the important case, when transversality of
the antisymmetric tensor field is preserved.  This conclusion is connected
with existence of the dual tensor $\tilde F^{\mu\nu}$ and with possibility
of the Bargmann-Wightman-Wigner-type quantum field theory revealed in
ref.~\cite{DVA0}. The remarkable feature of the Ahluwalia {\it et al.}
consideration is: boson and its antiboson can possess opposite relative
parities.

We choose the field operator, Eq. (\ref{fop}), in the following way:
\begin{mathletters}
\begin{eqnarray}
F^{i0}_{(+)} ({\bf p})
&=& E^i ({\bf p})\quad,\quad F^{jk}_{(+)} ({\bf p}) = - \epsilon^{jkl} B^l
({\bf p})\quad;\\ F^{i0}_{(-)} ({\bf p}) &=& \tilde F^{i0} = B^i ({\bf
p})\quad,\quad F^{jk}_{(-)} ({\bf p}) = \tilde F^{jk} = \epsilon^{jkl} E^l
({\bf p})\quad,
\end{eqnarray}
\end{mathletters}
where $\tilde F^{\mu\nu}
= {1\over 2} \epsilon^{\mu\nu\rho\sigma} F_{\rho\sigma}$ is the tensor
dual to $F^{\mu\nu}$; and $\epsilon^{\mu\nu\rho\sigma} = -
\epsilon_{\mu\nu\rho\sigma}$\, , \, $\epsilon^{0123} = 1$ is the totally
antisymmetric Levi-Civita tensor.
After lengthy but standard calculations we achieve:
\begin{eqnarray}
{\bf J}^k &=& \sum_{\eta\eta^\prime}\int \frac{d^3 {\bf p}}{(2\pi)^3 2E_p}
\left \{ \frac{i\epsilon^{ijk}{\bf E}^i_\eta ({ \bf p})
{\bf B}^j_{\eta^\prime} ({\bf p})}{2}
\left [ a_\eta ({\bf p}) b^\dagger_{\eta^\prime} ({\bf p}) +
a_{\eta^\prime} ({\bf p}) b^\dagger_\eta ({\bf p}) +
b^\dagger_{\eta^\prime} ({\bf p}) a_\eta ({\bf p}) +
b^\dagger_\eta ({\bf p}) a_{\eta^\prime} ({\bf p}) \right ] -
\right.\nonumber\\
&-&\left. \frac{i{\bf p}^k ({\bf E}_\eta ({\bf p}) \cdot {\bf
E}_{\eta^\prime} ({\bf p}) + {\bf B}_\eta ({\bf p}) \cdot {\bf
B}_{\eta^\prime})
- i {\bf E}^k_{\eta^\prime} ({\bf p}) ({\bf p}\cdot {\bf E}_{\eta} ({\bf
p}))
- i {\bf B}^k_{\eta^\prime} ({\bf p}) ({\bf p}\cdot {\bf B}_{\eta}
({\bf p}))}{2E_p}\times\right.\nonumber\\
&\times&\left.  \left [ a_\eta
({\bf p}) b^\dagger_{\eta^\prime} ({\bf p}) + b^\dagger_\eta ({\bf p})
a_{\eta^\prime} ({\bf p})\right ] \right \}
\end{eqnarray}
We should choose normalization conditions. For instance, one can use
an analogy with classical electrodynamics (a photon is massless):
\begin{mathletters}
\begin{eqnarray}
&&({\bf E}_\eta ({\bf p}) \cdot {\bf E}_{\eta^\prime} ({\bf p}) + {\bf
B}_\eta ({\bf p}) \cdot {\bf B}_{\eta^\prime})  = 2E_p
\delta_{\eta\eta^\prime}\quad,\\
&&{\bf E}_\eta \times {\bf B}_{\eta^\prime} = {\bf
p}\delta_{\eta\eta^\prime} -{\bf p}\delta_{\eta, -\eta^\prime}\quad.
\end{eqnarray}
\end{mathletters}
These conditions imply that ${\bf E}\perp {\bf B} \perp {\bf p}$.
Finally, we obtain
\begin{equation}
{\bf J}^k =  - i\sum_\eta \int \frac{d^3 {\bf p}}{(2\pi)^3}
\, \frac{{\bf p}^k}{2E_p} \left [ a_\eta b_{-\eta}^\dagger
+b_\eta^\dagger a_{-\eta} \right ]\quad.\label{spin}
\end{equation}
If we want to describe states  with definite helicity quantum number
(photons) we should assume that $b^\dagger_\eta ({\bf p})=
i a^\dagger_\eta ({\bf p})$ what is reminiscent with Majorana-like
theories~\cite{DVA1,DVA2,Bil}.\footnote{Of course, an imaginary unit
can be absorbed by the corresponding re-definition of negative-frequency
solutions.} One can take into account the prescription of the normal
ordering and set up the commutation relations in the form:
\begin{equation}
\left [a_\eta ({\bf p}), a_{\eta^\prime}^\dagger ({\bf k})\right
]_{-} = (2\pi)^3 \delta ({\bf p}-{\bf k}) \delta_{\eta,-\eta^\prime}\quad.
\label{cr}
\end{equation}
After acting the operator (\ref{spin}) on the
physical states, {\it i.e.}, $a_h^\dagger ({\bf p}) \vert 0>$ , we are
convinced that antisymmetric tensor field can describe particles with
transversal components (helicity is equal to $\pm 1$). One can see that
the origins of this conclusion are the possibility of different
definitions of the field operator and  the existence of `antiparticle' for
a particle described by antisymmetric tensor field. The latter statement
is related with the Weinberg discussion of the connection between
helicity and representations of the Lorentz group~[5a]. Next, I would
like to point out that the Proca-like equations for antisymmetric tensor
field with {\it mass}, {\it e.g.}, Eq. (\ref{PE}) can possess tachyonic
solutions, see for the discussion in ref.~\cite{DVO1}.  Therefore, in a
massive case the states can be ``partially" tachyonic ones. We then deal
with the problem of the choice of normalization conditions which could
permit us to describe both transversal and longitudinal modes of the $j=1$
field.

In conclusion, we calculated the Pauli-Lyuban'sky vector of relativistic
spin on the basis of the N\"otherian symmetry
method~\cite{Corson,Barut,Bogoliubov}.  Let me remind that it is a part of
the angular momentum vector, which is conserved as a consequence of
rotational invariance. After explicit~\cite{Hayashi} (or
implicit~\cite{AVD}) applications of the constraints (the generalized
Lorentz condition) in the Minkowski space, the antisymmetric tensor field
becomes longitudinal (helicity $h$ is equal to zero).  We proposed one of
possible ways to resolve this contradiction with the Correspondence
Principle in refs.~\cite{DVO1,DVO2,DVO3,DVO4}.  Another hypothesis has
been proposed by Evans~\cite{Evans,Evans1,EVANS}, in which the third
component of the Pauli-Lyuban'sky vector has been identified with the new
$B(3)$ field of electromagnetism.\footnote{See also the paper of
Chubykalo and Smirnov-Rueda~\cite{Chubykalo}. The paper on connections
between the Chubykalo and Smirnov-Rueda `action-at-a-distance' construct
and the $B(3)$ theory is in progress (private communication from A.
Chubykalo).} The present article continues these researches.  The
conclusion achieved is:  the antisymmetric tensor field can describe both
the Maxwellian $j=1$ field and the Kalb-Ramond $j=0$ field.  Nevertheless,
we still think that the physical nature of the $E=0$ solution revealed in
refs.~\cite{Gian,DVA00}, its connections with the Evans-Vigier $B(3)$
field, ref.~\cite{Evans,Evans1}, with Avdeev-Chizhov $\delta^\prime$- type
transversal solutions~[21b], which cannot be interpreted as relativistic
particles, as well as with my concept of $\chi$ boundary functions,
ref.~\cite{DVO4} are not completely explained until now.   Finally, while
we do not have any intention to doubt theoretical results of the ordinary
quantum electrodynamics (it deals mainly with $E=\pm p$ solutions), we
are sure that questions put forth in this note (as well as in previous
papers of both mine and other groups) should be explained properly.

\medskip
{\it Acknowledgements.}
I am thankful to Profs. D. V. Ahluwalia, A. E. Chubykalo,
A. F. Pashkov and S. Roy for stimulating discussions. After the writing of
the preliminary version of the manuscript I received the paper of
Professor M.  W. Evans ``The Photomagneton and Photon
Helicity"~\cite{EVANS}, devoted to a consideration of the similar
topics, but from different points.  His patient elucidations of the
Evans-Vigier $B(3)$ model and useful information are acknowledged.
I am delighted by the quality of referee reports on
the papers~\cite{DVO1,DVO2,DVO3,DVO4} from ``Journal of Physics A".
In fact, they helped me to learn many useful things. The
writing of the present paper has been inspired by the book of
M.  Ancharov ``Kak ptitza Garuda".

I am grateful to Zacatecas University,  M\'exico, for a Full Professor
position. This work has been partially supported by the Mexican Sistema
Nacional de Investigadores and the Programa de Apoyo a la Carrera Docente.

\end{document}